\documentclass[useAMS,usenatbib]{mnras}
\usepackage{graphicx}
\usepackage{epstopdf}
\usepackage{amsmath}
\usepackage{amssymb}
\usepackage{bm}		% Bold maths symbols, including upright Greek
\usepackage{booktabs,caption}
\usepackage[flushleft]{threeparttable}

\title[Radiation-driven outflow in AGNs]{Radiation-driven outflows in AGNs:
Revisiting feedback effects of scattered and reprocessed photons}
\author[A. Mosallanezhad et al.]
{Amin Mosallanezhad$^{1,2}$\thanks{E-mail: mosallanezhad@ustc.edu.cn}, Feng Yuan$^{2,3}$\thanks{E-mail: fyuan@shao.ac.cn}, Jeremiah P. Ostriker$^{4}$,
\newauthor
Fatemeh Zahra Zeraatgari$^{5}$, and  De-Fu Bu$^{2}$ \\
$^{1}$ Key Laboratory for Research in Galaxies and Cosmology, Department of Astronomy, University of Science and Technology of China, \\ Hefei, Anhui 230036, PR China  \\
$^{2}$ Key Laboratory for Research in Galaxies and Cosmology, Shanghai Astronomical Observatory, Chinese Academy of Sciences, \\ 80 Nandan Road, Shanghai 200030, PR China \\
$^{3}$ School of Astronomy and Space Sciences, University of Chinese Academy of Sciences, No. 19A Yuquan Road, Beijing 100049, China \\
$^{4}$ Department of Astronomy, Columbia University, 550 W, 120th Street, New York, NY10027, USA\\
$^{5}$ School of Mathematics and Statistics, Xi’an Jiaotong University, Xi’an 710049, China}
\begin{document}

%\date{Accepted 1988 December 15. Received 1988 December 14; in original form 1988 October 11}

\pagerange{\pageref{firstpage}--\pageref{lastpage}} \pubyear{2002}

\maketitle

\label{firstpage}
\begin{abstract}
We perform two-dimensional hydrodynamical simulations of slowly rotating accretion flows in the region of $ 0.01-7\, \mathrm{pc} $ around a supermassive black holes with $ M_\mathrm{BH} = 10^{8} M_{\odot} $. The accretion flow is irradiated by the photons from the central active galactic nucleus (AGN). In addition to the direct radiation from the AGN, we have also included the ``re-radiation'', i.e., the locally produced radiation by Thomson scattering, line  and bremsstrahlung radiation. Compare to our previous work, we have improved the calculation of radiation force due to the Thomson scattering of X-ray photons from the central AGN. We find that this improvement can significantly increase the  mass flux and velocity of  outflow. We have compared the properties of outflow --- including mass outflow rate, velocity, and kinetic luminosity of outflow --- in our simulation with the observed properties of outflow in AGNs and found that they are in good consistency. This implies that the combination of line and re-radiation forces is the possible origin of observed outflow in luminous AGNs.

\end{abstract}


\section{Introduction}

It is believed that AGN feedback plays
an important role in the formation and evolution of their host galaxies
(\citealt{Fabian 2012}; \citealt{Kormendy and Ho 2013}). The most remarkable
observational evidence is the strong correlation between the mass of the
supermassive black holes (SMBHs) and the properties of the host galactic
bulge, such as luminosity (\citealt{Kormendy and Richstone 1995}), velocity
dispersion (\citealt{Gebhardt et al. 2000}; \citealt{Ferrarese and Merritt 2000};
\citealt{Tremaine et al. 2002}; \citealt{Gultekin et al. 2009};
\citealt{Graham et al. 2011}), and stellar mass (\citealt{Magorrian et al. 1998};
\citealt{Marconi and Hunt 2003}; \citealt{Haring and Rix 2004}).
The importance of AGN feedback  has also been confirmed
by many theoretical studies especially numerical simulations, from cosmological 
to galactic scales (see e.g., \citealt{Ostriker et al. 2010};
\citealt{Gan et al. 2014, Gan et al. 2019}; 
\citealt{Yuan et al. 2018}; 
\citealt{Yoon et al. 2018}; 
\citealt{Li et al. 2018}).

The three kinds of outputs from  AGNs  are radiation, jet, and wind. Especially, it has been found that wind plays an important role in AGN feedback (e.g., \citealt{Ostriker et al. 2010}). When the accretion rate of AGN is low, the accretion and feedback are in the hot mode. In this case, wind is driven by the combination of thermal and magnetic forces and the mass flux of wind can be much larger than the net accretion rate (e.g., \citealt{Blandford and Begelman}; \citealt{Yuan et al. 2012a, Yuan et al. 2012b}; \citealt{Narayan et al. 2012}; \citealt{Li et al. 2013}; \citealt{Yuan et al. 2015}; \citealt{Bu et al. 2016}; \citealt{Mosallanezhad et al. 2016}). This kind of ``hot'' wind has been confirmed by observations (e.g., \citealt{Wang et al. 2013}; \citealt{Tombesi et al.
2014}; \citealt{Homan et al. 2016}; \citealt{Cheung et al. 2016}; \citealt{Ma et al. 2019}) and found to play an important role in AGN feedback (e.g., \citealt{Weinberger et al. 2017a}; \citealt{Yuan et al. 2018}).

In the case of the cold accretion mode, a wind can be driven by the AGN radiation and magnetic field of the accretion disc.  In this paper we will focus on the radiation mechanism. If the gas is fully ionized,  radiation exerts a force on the gas through Thomson scattering. In the case of partially ionized gas,  the cross section of interaction between UV photons and the gas can be several thousands of times larger than that of Thomson scattering (e.g., \citealt{Proga et al. 2000}; \citealt{Proga 2007a}; \citealt{Proga et al. 2008}; \citealt{Korusawa and Proga 2008, Korusawa and Proga 2009}). Therefore, line force can be significantly stronger than the radiation force due to Thomson scattering.

Following this line, \citet{Korusawa and Proga 2009} (hereafter KP09) have presented
hydrodynamical simulations to investigate the interaction between radiation and slowly rotating gas
at parsec (pc) scale. Their computational domain covers a range from $ \sim  10^{-2} $ to $ 7 $ pc. They assumed that when gas falls into the region $ < 10^{-2} $ pc, an accretion disc will be formed. A spherical hot corona is assumed to be present above the disc and is responsible for the X-ray radiation of the AGN. The total accretion luminosity is self-consistently determined by calculating mass accretion rate at the inner boundary of the computational domain and by assuming a canonical radiative efficiency of the standard thin disc.  The  X-rays radiation will ionize the accretion flow at pc scale, and  will also produce a radiation pressure through Thomson scattering. For the optical/UV photons from the AGN, in addition to the  radiation force due to Thomson scattering, they will also interact with partially ionized gas and exert a much stronger line force.

In KP09 and other similar previous works, the authors only considered the radiation directly  coming from the central AGNs. In reality, the locally generated photons  through scattering, bremsstrahlung and line radiation in the gas far away from the AGN can also play an important role. The radiation force due to the locally generated photons is called ``re-radiation'' force.
\citet{Liu et al. 2013} investigated the effects of re-radiation force on the dynamics of the flow at parsec scale and found that  the accretion flow becomes thicker due to the additional vertical component of the re-radiation force and the outflow becomes stronger consequently.

For calculating the radiation force due to the Thomson scattering of X-ray photons from the central AGN, KP09 made use of $ F=F_{X} \kappa_{es}/c $ ($ F_{X} $ is the X-ray flux, $ \kappa_{es} $ is Thomson scattering opacity, $ c $ is the speed of light).
However in \citet{Liu et al. 2013}, the author used another formula to calculate this force. The formula adopted in \citet{Liu et al. 2013} is that $ F = \dot{E} / (\rho c) $, with $ \dot{E} $ being the Compton heating rate, and $ \rho $ is the gas density. We think the formula used in \citet{Liu et al. 2013} does not correctly represent the force due to the Thomson scattering. For instance, this force will be incorrectly equal to zero if the ``Compton temperature'' is equal to the gas temperature. Therefore, in the present paper, we use the same formula adopted in KP09 to calculate the radiation force due to the Thomson scattering of X-ray photons from the central AGN. We will make this correction and investigate how the result will be changed compared to \citet{Liu et al. 2013}.

There is some observational evidence for AGNs with sub-Eddington luminosity. For instance, \citealt{Kollmeier et al. 2006} in a study on 407 AGNs in the redshift range $ Z \sim 0.3-4 $ has found that the most of the AGNs in the sample are sub-Eddington. Indeed, they described the luminosity distribution of the estimated Eddington ratios by log-normal which the maximum value is $L_{\rm bol}/L_{\rm Edd} \approx 1/4$ ($L_{\rm bol}$ and $L_{\rm Edd}$ being bolometric and Eddington luminosity, respectively. \citealt{Steinhardt and Elvis 2010} used a large sample of 62185 quasars from SDSS. They also obtained same results. In fact, the broad mass distribution of the fuelling gas of the galactic scale can provide enough gas to fuel the AGNs to have super-Eddington luminosities. Moreover, based on slim accretion disk model, the rotating accretion disk can have super-Eddington luminosity (\citealt{Abramowicz et al. 1988}; \citealt{Ohsuga et al. 2005}; \citealt{Yang et al. 2014}). Therefore, the sub-Eddington accretion of the AGNs is a problem which should be solved.
 
One proposed solution for the problem of sub-Eddington puzzle is AGN radiative feedback. Nevertheless, the inefficiency of the radiative feedback for innermost region of the slim disk is shown by numerical simulation done by \citealt{Ohsuga et al. 2005}. They showed, by considering radiative transfer, the central region of the black hole accretion flow can radiate well above the Eddington luminosity. On parsec scale, KP09 showed that the line force can drive strong wind, however the radiative feedback could not solve the sub-Eddington puzzle in their numerical results. In this paper, we try to solve this puzzle by adding the re-radiation force.

%In \citet{Liu et al. 2013}, when they calculate the radiation force due to the direct radiation from the central AGN, the force corresponding to the  Thomson scattering between AGN X-ray photons  and accretion gas is missing. Or more precisely, although they include one term describing the interaction between X-ray and the gas (the $G_{\rm Comp}$ term in the $\dot{E}_{\rm Comp}$ term in equation (13) of Liu et al. 2013), this term does not correctly represent the force due to the Thomson scattering. For instance, this term will be incorrectly equal to  zero if the ``Compton temperature'' is equal to the gas temperature.  Since this is potentially an important term, in the present work, we will make a correction and investigate how the result will be changed compared to Liu et al. (2013).

The structure of the paper is as follows. We describe our method, model assumptions, and modifications to the \citealt{Liu et al. 2013} work in sections \ref{sec:Numerical_Method} and \ref{sec:Radiation}. The results of our simulations will be given in Section \ref{sec:Results}. In section \ref{sec:Summary_Discussion}, we will discuss and summarize the main results.

\section{Numerical Method} \label{sec:Numerical_Method}

We perform axisymmetric two-dimensional hydrodynamic (HD) simulations, by using
grid-based multidimensional  code \verb'ZEUS-MP'
(\citealt{Hayes et al. 2006}), which is the massive parallel MPI-implemented version
of the \verb'ZEUS-3D' code (\citealt{Hardee and Clarke 1992}; \citealt{Clarke 1996}).
Our basic physical setup is mostly the same as one used in \citealt{Liu et al. 2013} (see also \citealt{Korusawa and Proga 2008, Korusawa and Proga 2009}).
We outline our
numerical method and all differences from \citealt{Liu et al. 2013} below.

\subsection{Basic Equations}

We consider a SMBH located at the origin of the polar coordinate
system which is surrounded by an accretion flow. To compute the evolution
and the structure of the accretion flow irradiated by the strong radiation from AGN,
we solve the following set of equations,

\begin{equation} \label{continuity}
  \frac{d\rho}{dt} + \rho \nabla \cdot \bm{v} = 0,
\end{equation}
\begin{equation} \label{momentum}
  \rho \frac{d \bm{v}}{d t} = - \nabla p - \rho \nabla \psi + \rho \bm{F}_\mathrm{rad},
\end{equation}
\begin{equation} \label{energy}
  \rho \frac{d (e/\rho)}{dt} = -p \nabla \cdot \bm{v} + \rho \mathcal{L},
\end{equation}
where $ \rho $, $ \bm{v} $, $ p $, $ e $, and $ \psi $ are mass density, velocity, gas pressure,
internal energy density, and gravitational potential, respectively. We employ the pseudo-Newtonian
potential, $ \psi = -GM/(r - r_\textrm{s}) $ (\citealt{Paczynsky and Wiita 1980}), where $ M $ and $ G $ are
the centre BH mass and the gravitational constant, respectively, and $ r_\textrm{s} \equiv 2GM/ c^{2} $.
The Lagrangian/comoving derivative is given by $ d/dt \equiv \partial/\partial t + \bm{v} \cdot \nabla $.
We adopt the adiabatic equation of state in the form of $ p = (\gamma - 1) e $, where $ \gamma $ is
the adiabatic index and is set to be $ \gamma = 5/3 $. Here, $ \bm{F}_\mathrm{rad} $ is the total radiation
force per unit mass, and $ \mathcal{L} $ in the energy equation, Equation (\ref{energy}), denotes the net cooling rate.
Both radiation force and net cooling rate will be described in more details in the following section.

\subsection{Model Setup}

To solve equations (\ref{continuity})-(\ref{energy}), we use spherical polar coordinates $ (r, \theta, \phi) $,
where $ r $ is the distance from the origin of the coordinates, $ \theta $ is the polar angle and
$ \phi $ is the azimuthal angle. %Moreover, the axial symmetry about the rotation axis of the accretion
%disc ($  \theta = 0^{\circ} $) is assumed.
%
We set the two-dimensional computational domain of $ r_\mathrm{min} \leq r \leq  r_\mathrm{max} $ and
$ \epsilon \leq \theta \leq \pi / 2  $, where $ \epsilon $ is set to be a small value to
avoid the numerical singularity near the polar axis. For all the runs presented here, we set
$ r_\mathrm{min} =  500 r_{*} $, and $ r_\mathrm{max} = 2.5 \times10^{5} r_{*} $, where
$ r_{*} = 3 r_\mathrm{s} $ is the innermost stable circular orbit (ISCO) of a Schwarzschild BH.
We divide the $ r-\theta $ plane into zones as follows: in $r$ direction, we have 144 zones with the zone size ratio $ dr_\text{i+1}/dr_\text{i} = 1.04 $, which ensures good resolution near the inner region of our computational domain.
In the $ \theta $ direction, we have 64 zones with
$ d \theta_\text{j+1}/d \theta_\text{j} = 1.0 $ (i.e., equally spaced grids).

For initial conditions, we adopt the uniform density and gas temperature everywhere
in the computational domain, i.e., $  \rho(r,\theta) = \rho_{0} $ and $ T(r, \theta) = T_{0} $.
We set the initial radial and latitudinal components of the velocity to be zero,
$ v_{r}(r, \theta) = v_{\theta}(r, \theta) = 0 $, and the angular velocity of the gas is
assigned to have the following specific angular momentum distribution,
\begin{equation}\label{initial_vp}
  v_{\phi}(r, \theta) =
\begin{cases}
0 \quad \quad \quad &\text{for} \ r < 10^{5} r_{*}\\
l/( r \sin\theta )  &\text{for} \ r \geq 10^{5} r_{*}.
\end{cases}
\end{equation}
where, $ l $ is the latitude-dependent specific angular momentum given as,
\begin{equation}\label{angularmom}
  l(\theta) = l_{0} (1 - \left| \cos \theta \right| ), \quad \quad l_{0} = \sqrt{GM_\mathrm{BH} r_\mathrm{cir}} .
\end{equation}
Here, $ r_\mathrm{cir} $ is the ``circularization radius" on the equatorial plane.
We set $ r_\mathrm{cir} = 300 r_{*} $, which is much smaller than the radial inner boundary of our simulation domain.
We consider the relatively small value of $ r_\mathrm{cir} $ to prevent the formation of a rotationally supported torus in our computational domain and the complexities associated with it.

The boundary conditions are set as follows.
We apply axisymmetric  boundary conditions at the rotation axis (i.e., $ \theta = 0 $) and
reflecting boundary conditions at the equatorial plane ($ \theta = \pi / 2 $).
At the inner radial boundary, we use outflow boundary conditions (e.g., \citealt{Stone and Norman 1992}). At the outer radial boundary, if the gas flows in ($ v_{r}(r_\mathrm{max}, \theta) < 0 $), all HD
quantities except the radial component of the velocity, $ v_{r} $, are set to the
initial conditions, i.e., $ \rho(r_\mathrm{max}, \theta) = \rho_{0} $,
$ T(r_\mathrm{max},\theta) = T_{0} $, $ v_{\theta}(r_\mathrm{max},\theta) = 0 $, and
$ v_{\phi} (r_\mathrm{max},\theta) = l/(r_\mathrm{max} \sin\theta ) $.  When $ v_{r}(r_\mathrm{max}, \theta) > 0 $, we use outflow boundary conditions. This approach is adopted to mimic
the situation where there is always gas available for accretion and represents steady
conditions at the outer radial boundary.

\section{Radiative heating/cooling and radiation force} \label{sec:Radiation}

\subsection{Modeling of the central AGN}

We assume that the accretion flow in the central AGN  has
two components: a standard thin disc located at the equator (\citealt{Shakura and Sunyaev 1973}), and a hot corona  above the thin disc. We also assume that the innermost edge of the disc is located at $r_{*}$ and the outer boundary of the disc is much
smaller than the inner boundary of the computational domain, i.e., $ r_\mathrm{d} \ll r_\mathrm{min} $.
We assume that the disc only emits optical/UV photons (i.e., $ L_\mathrm{d}  = L_\mathrm{UV} $),
and the hot corona only emits X-ray, $ L_{*} = L_\mathrm{X} $. The total luminosity then includes both the luminosity of the disc and the corona, $L _\mathrm{acc} =L_\mathrm{d}+L_{*}$. Following KP09, we further assume that the extended disk radiation contributes to the radiation force due to both line force as well as electron scattering. Whereas, the corona radiation contributes to the radiation force only due to the electron scattering (line force contributions of the corona radiation has been ignored). We define the
parameter $ f_{*} $ as the ratio of the corona luminosity to the total luminosity,
$ L_{*} = f_{*} L_\mathrm{acc} $, and  $ f_\mathrm{d}  = 1 - f_{*} $
as the fraction of the total luminosity in the disc emission. We set $ f_{*}  = 0.05 $
throughout this paper.
%We suppose that the corona is contributed to the radiation force only due
%to electron scattering and responsible for ionizing the gas to the high ionization state.
%Therefore, the disc radiation force is due to both electron scattering as well as
%line force.

Since the inner boundary of our computational domain is far from the outer edge of the disc,
$ r_\mathrm{d} \ll r_\mathrm{min} $,   we can safely
adopt point-source approximation,
\begin{equation} \label{coronaflux}
   \mathcal{F}_{*}(r) = \frac{L_{*}}{4 \pi r^{2}} = \frac{f_{*} L_\mathrm{acc}}{4 \pi r^{2}},
\end{equation}
\begin{equation} \label{discflux}
    \mathcal{F}_\mathrm{d}(r, \theta) = 2 \left| \cos\theta \right|  \, \frac{f_\mathrm{d} L_\mathrm{acc}}{4 \pi r^{2}}.
\end{equation}
In equation (\ref{discflux}), the factor ``2'' comes from the normalization of the flux.
In the present work, we do not consider the UV attenuation due to Thomson scattering, i.e., we assume $ \tau_\mathrm{x} = \tau_\mathrm{uv} =0$ in the calculations of radiation fluxes from the central engine. This is based on two considerations. The first reason is as stated in detail in KP09 and references therein. The calculation of line force adopted here (and also in other works such as KP09 and \citealt{Liu et al. 2013}) is taken from a simplified approach proposed in \citealt{Stevens and Kallman 1990}. But this approach is based on the assumption that the radiation is optically thin. So to be consistent, we neglect the optical depth\footnote{We do include the X-ray optical depth when we calculate the ionization parameter}. In our parameter regime, optical depth is dominated by scattering. Unlike
true absorption, scattering merely re-directs the photons. The scattered
X-ray photons must be replenished by photons scattered from
other propagation lines. Indeed, with pure scattering there is no attenuation, but only re-direction.
%It is hard to precisely calculate the real attenuation so we simply assume another extreme in the present work, i.e, scattering is neglected. 
This is another reason why we neglect the optical depth.
%$ \tau_\mathrm{x} $ will be only used for evaluating the photoionization parameter. This is our first
%modification to \citealt{Liu et al. 2013} work. \footnote{We do make this change since the formula for
%calculating the line force multiplier only applies to UV optically thin case (cf. the related discussion in %KP09).}

\subsection{Mass accretion rate and accretion luminosity}

In our simulation, at each time step, the accretion luminosity will be updated self-consistently based on
the  mass accretion rate, $ \dot{M}_\mathrm{a} (t) $, calculated at the inner boundary of our
computational domain, $ r_\mathrm{min} $. The accretion luminosity is then,
\begin{equation} \label{Lacc}
   L_\mathrm{acc} (t) =  \eta \dot{M}_\mathrm{a} (t) c^{2},
\end{equation}
where $ \eta $ is the radiative efficiency and set to be $ \eta = 1/12  $ in this paper since we assume the disc is described by the standard thin disc model.
Following KP09 and \citealt{Liu et al. 2013}, the time averaged mass accretion rate can be expressed as,
\begin{equation} \label{accretionrate}
	\dot{M} _\mathrm{a}(t) = \frac{\int^{t-\tau}_{t-\tau-\Delta t} \dot{M}(t^{\prime}) dt^{\prime}}{\int^{t-\tau}_{t-\tau-\Delta t}  dt^{\prime}},
\end{equation}
where $ \tau $ denotes a time lag between the change of the mass accretion rate and
the change of the accretion luminosity from the central engine. Since the standard
thin disc model is considered here, the lag time can be approximated by accretion
time scale as $ t_\mathrm{acc} \approx r_\mathrm{d}/ v_\text{r} = r_\mathrm{d}/(\alpha c_\mathrm{s} H / r_\mathrm{d} )  $.
Here, $ c_\mathrm{s} $ is the isothermal sound speed, $ H $ is the half thickness of
the disc and  $ \alpha $ is the viscosity parameter. According to our settings, the
lag time will be approximately $ \tau \approx 10^{9} $ seconds\footnote{It should be noted here that the
results are not so sensitive to the value of $ \tau $.}. By setting $  \Delta t = \tau $
the denominator of the equation (\ref{accretionrate}) becomes $ \tau $.

\subsection{Radiative cooling/heating} \label{subsec:Radiative_cooling}

The net cooling rate in the energy equation consists of four heating/cooling terms
including Compton heating/cooling, $( G_{\mathrm{Comp}} ) $, X-ray photoionization
heating and recombination cooling, $ (G_\mathrm{X}) $, cooling via line emission
$ (L_\mathrm{line}) $  and bremsstrahlung cooling $ (L_\mathrm{brem}) $.
The net cooling rate is as follows (\citealt{Blondin 1994}):
\begin{equation} \label{netrate}
     \rho \mathcal{L} = n^{2}  \left( G_{\mathrm{Comp}} + G_\mathrm{X} - L_\mathrm{line}  - L_\mathrm{brem} \right),
\end{equation}
where
\begin{equation} \label{Gcomp}
     G_{\mathrm{Comp}}  =  8.9 \times 10^{-36} \xi \left(T_{X}  -  4 T \right),
\end{equation}
\begin{equation} \label{GX}
     G_\mathrm{X}  =1.5 \times 10^{-21} \xi^{1/4} T^{-1/2} \left(1 - T/ T_{X} \right),
\end{equation}
\begin{equation} \label{Lline}
L_\mathrm{line} = \left[ 1.7 \times 10^{-18} \exp (T_{l}/T ) \xi^{-1} T^{-1/2} + 10^{-24} \right] \delta,
 \end{equation}
\begin{equation} \label{Lbrem}
L_\mathrm{brem} = 3.3 \times 10^{-27} T^{1/2},
\end{equation}
where $ n = \rho/ (\mu m_{p}) $ is the number density of the local gas located at the distance
$ r $ from the AGN, $ m_{p} $ denotes the proton mass, and  $ \mu $ is the mean molecular
weight which is set to be $ 1 $. In the above equations,
$ T_\mathrm{X} = 8 \times 10^{7} \, \text{K} $ is the ``characteristic temperature'' or ``Compton'' temperature
of the X-ray radiation (\citealt{Sazonov et al. 2004}) and  $ T_{l} $ parameterizes the line cooling
temperature $ (T_{l} = 1.3 \times 10^{5} \, \text{K} ) $. The parameter $ \delta $ in the line cooling rate is
used to control line cooling ($ \delta < 1 $ represents optically thick cooling and $ \delta = 1 $ represents
optically thin cooling). We set $ \delta = 1 $. In general, the line cooling dominates
over the other cooling processes. According to the Blondin formula, the net cooling rate depends on the
gas temperature,  $ T $,  density, and the photoionization parameter $ \xi $. The photoionization parameter
is expressed as,
\begin{equation} \label{ksi}
    \xi  \equiv \frac{f_{*} L_\text{acc} }{n r^{2}} \mathrm{e}^{-\tau_\mathrm{x}},
\end{equation}
where,
\begin{equation} \label{taux}
    \tau_\mathrm{x} = \int_{r_\mathrm{min}}^{r} \rho \kappa_\mathrm{x} \rm{d}r.
\end{equation}
Here, $ \tau_\mathrm{x} $ is the X-ray scattering optical depth in
the radial direction, and $ \kappa_\mathrm{x} =  0.4 \, \mathrm{cm}^{2} \mathrm{g}^{-1} $ is the opacity.

\subsection{The radiation force} \label{subsec:radiation_force}

Both the photons from the central AGN and those locally produced  can exert radiation force on gas, so
\begin{equation} \label{total force}
    \bm{F}_\mathrm{rad} = \bm{F}_\mathrm{c}  + \bm{F}_\mathrm{re},
\end{equation}
where $ \bm{F}_\mathrm{c} $ denotes the radiation force due to the central AGN and $ \bm{F}_\mathrm{re} $
denotes the re-radiation force due to the locally produced photons.

Let us first calculate $\bm{F}_\mathrm{c}$.  First,  since the gas is not fully ionized, there should be a force corresponding to photoionization heating-recombination cooling (e.g., \citealt{Liu et al. 2013} and references therein):
\begin{equation} \label{Edot}
   \dot{E}_\mathrm{X}/\rho c = n^{2} G_\mathrm{X}/\rho c.
\end{equation}

For the optical/UV radiation from the AGN, there are two components of force, which are due to  Thomson scattering and line force, respectively,
\begin{equation}
\bm{F}_{\rm{c,o}}=2\frac{\kappa_{es}}{c}\frac{L_{\rm acc}}{4\pi r^2}f_d[1+\mathcal{M}(t)]\left| \cos\theta \right| \hat{\bm{r}}.
\end{equation}
In the above equation, $ \mathcal{M} $ represents the force multiplier
(\citealt{Castor et al. 1975}) that is to parameterize how much spectral lines
increase the scattering coefficient (see \citealt{Proga et al. 2000} for more details).

For the X-ray radiation from the AGN, we  only consider the force due to Thomson scattering:
\begin{equation}
\bm{F}_{\rm{c,X}}=\frac{\kappa_{es}}{c}\frac{L_{\rm acc}}{4\pi r^2}f_*\hat{\bm{r}}.
\end{equation}

The total force due to radiation from the central AGN is,
\begin{equation} \label{F_cent}
\bm{F}_\mathrm{c}=\frac{n^{2} G_\mathrm{X}}{\rho c} + \bm{F}_{\rm{c,o}} + \bm{F}_{\rm{c,X}}.
\end{equation}
Compare this equation with equation (13) in \citealt{Liu et al. 2013},  the difference is that we now correctly include the force due to Thomson scattering of X-ray photons from the central AGN.  We expect that the wind will become stronger as a consequence.

%To evaluate the radiative acceleration due to the disc line force, we assume that only
%optical/UV photons have contributions. To calculate line force, we generally follow \citealt{Castor et al. 1975}.
%The line force at a point $ r $ can be described as,
%
%\begin{equation}\label{line_force}
%	\bm{F}_\mathrm{line} = \oint_{\Omega}  \mathcal{M}(t) \left[ \frac{\kappa_\mathrm{es} %I(\bm{r},\hat{\bm{n}})}{c} \right]  \hat{\bm{n}} \  \mathrm{d} \Omega,
%\end{equation}
%
%where $ I $, $ \Omega $ are the frequency integrated continuum
%intensity in the direction $ \hat{\bm{n}} $, and the solid angle subtended by the source of
%radiation, respectively.
%The radiation force due to photons from central AGN is described in Equations (18)-(21). The total radiation force from the central AGN can be written as,
%\begin{equation} \label{F_c_total}
%    \bm{F}_\mathrm{c} =  \frac{n^2G_{\rm X}}{\rho c}  + \frac{\kappa_\mathrm{es}}{c}\frac{  L_\mathrm{acc} }{ 4 \pi r^{2} }  \left( f_{*} + 2 f_\text{d} \left[  1 + \mathcal{M}(t) \right] \left| \cos \theta \right| \right) \ \hat{\bm{r}}.
%\end{equation}
%Note that compared to Liu et al. (2013), in the first term of the right-hand side of the above equation, we %have removed the term corresponding to the Compton scattering because it has been included in the %second term of the right-hand side.

The calculation of the re-radiation force (the second term in equation (\ref{total force})) is exactly the same as that in \citealt{Liu et al. 2013}. For convenience, we briefly introduce as follows. The plane-parallel approximation is adopted in our simple radiative transfer calculation.
In this case, the re-radiation force will be in z-direction. By applying the Gauss theorem,
the vertical radiative force is described by,
\begin{equation} \label{fradz}
   \bm{F}_\mathrm{re} = \frac{\kappa_\mathrm{es}}{c} \int^{z}_{0} \left(  S_\mathrm{c} + n^2 L_\mathrm{brem} + n^2 L_\mathrm{line} \right) \mathrm{d} z
\end{equation}
where
\begin{equation} \label{sourceterm}
    S_\mathrm{c} = \rho \kappa_\mathrm{es} \cdot \frac{L_\mathrm{acc}}{ 4 \pi r^{2}} \left[  f_{*} + 2\, f_{d} \left| \cos \theta \right|   \right]
\end{equation}
is the source term due to the first-order scattered photons of the radiation from the central AGN.

\section{Results} \label{sec:Results}

\begin{table*}
\begin{center}
\caption{Summary of models with different parameters.} % title of Table
\label{table1} % is used to refer this table in the text
\begin{tabular}{c c c c c c c} % centered columns (4 columns)
%\begin{tabular}{lllllll}
\hline %inserts double horizontal lines
\hline
Run & Model & $ \rho_{0} $ & $ T_{0} $ & $ L_\mathrm{acc} $ & $ \dot{M}_\mathrm{out}(r_\mathrm{max}) $ & $ \eta_{w} $  \\  [0.5ex] % inserts table
    Number &   &  ($10^{-21} $\, g \,cm$^{-3}$) & ($ 10^{6} $\, K) & ($ L_\mathrm{Edd} $) & ($ 10^{25}$\, g\, s$^{-1} $) &  \\ [0.5ex] % inserts table
(1)   & (2) & (3) & (4) & (5) & (6) & (7) \\ [0.5ex] % inserts table
%\heading
\hline % inserts single horizontal line
1   &  R5a  &  10   &  2 & 0.77(0.04) & 37.95(8.71)   &  2.88(0.70) \\ % inserting body of the table
2   &  R6a  &  20   &  2 & 1.19(0.16) & 67.30(10.90) &  3.35(0.78) \\ % inserting body of the table
3   &  R7a  &  50   &  2 & 2.02(0.02) & 139.13(5.03) &  3.99(0.14) \\ % inserting body of the table
4   &  R8b  &  100 &  2 & 3.14(0.27) & 202.05(28.40)& 3.75(0.59) \\ % inserting body of the table
...  &    ...    & ...    & ... & ... & ... & ... \\
5   &  M5    & 10   &  2 & 0.79(0.1) & 46.84(6.5) &  3.62(1.33) \\
6   &  M6    & 20   &  2 & 1.01(0.02) & 82.46(16.3) & 4.89(0.35) \\
7   &  M7    & 50   &  2 & 1.64(0.08) & 134.12(29.75) & 4.93(0.4) \\
8   &  M8    & 100 &  2 & 1.97(0.46) & 404.05(38.13) & 12.44(1.51) \\
 ... & ...       & ...    & ... & ... & ... \\
9   &  M9    & 10   &  20 & 1.29 (0.01) & 123.14(8.65) & 5.71(0.78) \\
10 &  M10  & 50   &  20 & 2.61(0.07) & 1000.48(16.64) & 23.15(3.34) \\
\hline %inserts single line
\end{tabular}
\begin{tablenotes}
      \item \textit{Note}. Values in parentheses in Columns 5, 6 and 7 are the normalized standard deviations $ \sigma_{n} $ of the time series values. 
    \end{tablenotes}
\end{center}
\end{table*}

We have performed 10 simulations with different values  of $ \rho_0 $ and $ T_0 $  at the outer boundary to
explore the effects of these parameters (see columns 3 and 4 of Table \ref{table1}).
The model parameters and some basic results are summarized in Table \ref{table1}.
In models 1, 2, 3 and 4, the calculation of radiation force due to central AGN X-ray photons is the same as that in \citealt{Liu et al. 2013}. In models 5-10, we use the new formula to calculate the radiation force due to central AGN X-ray photons (see Section \ref{subsec:radiation_force}). To quantitatively study the flows, we calculate the mass inflow, outflow, and net rate as follows,

\begin{equation} \label{inflow rate}
     \dot{M}_\mathrm{in}(r) = 4 \pi r^{2} \int^{\pi/2}_{0} \rho\, \mathrm{min}(v_{r},0)\, \sin(\theta)\, \mathrm{d} \theta,
\end{equation}
\begin{equation} \label{outflow rate}
     \dot{M}_\mathrm{out}(r) = 4 \pi r^{2} \int^{\pi/2}_{0} \rho\, \mathrm{max}(v_{r},0)\, \sin(\theta)\, \mathrm{d} \theta,
\end{equation}
\begin{equation} \label{net rate}
     \dot{M}_\mathrm{net}(r) = 4 \pi r^{2} \int^{\pi/2}_{0} \rho\, v_{r} \, \sin(\theta)\, \mathrm{d} \theta.
\end{equation}
The time-averaged accretion luminosity and mass outflow rate
are given in columns 5 and 6 of Table \ref{table1}, respectively. The last column of
Table \ref{table1} gives the ratio of mass outflow rate
at the outer boundary $ (r_\mathrm{max}) $ to the mass accretion
rate at the inner boundary $ (r_{\min}) $, i.e., $  \eta_{w}
= \dot{M}_\mathrm{out} (r_\mathrm{max}) / \dot{M}_\mathrm{a} (r_\mathrm{min}) $.

\subsection{The fiducial run} \label{the_fiducial_run}

\begin{figure*}
 \includegraphics[width=\textwidth]{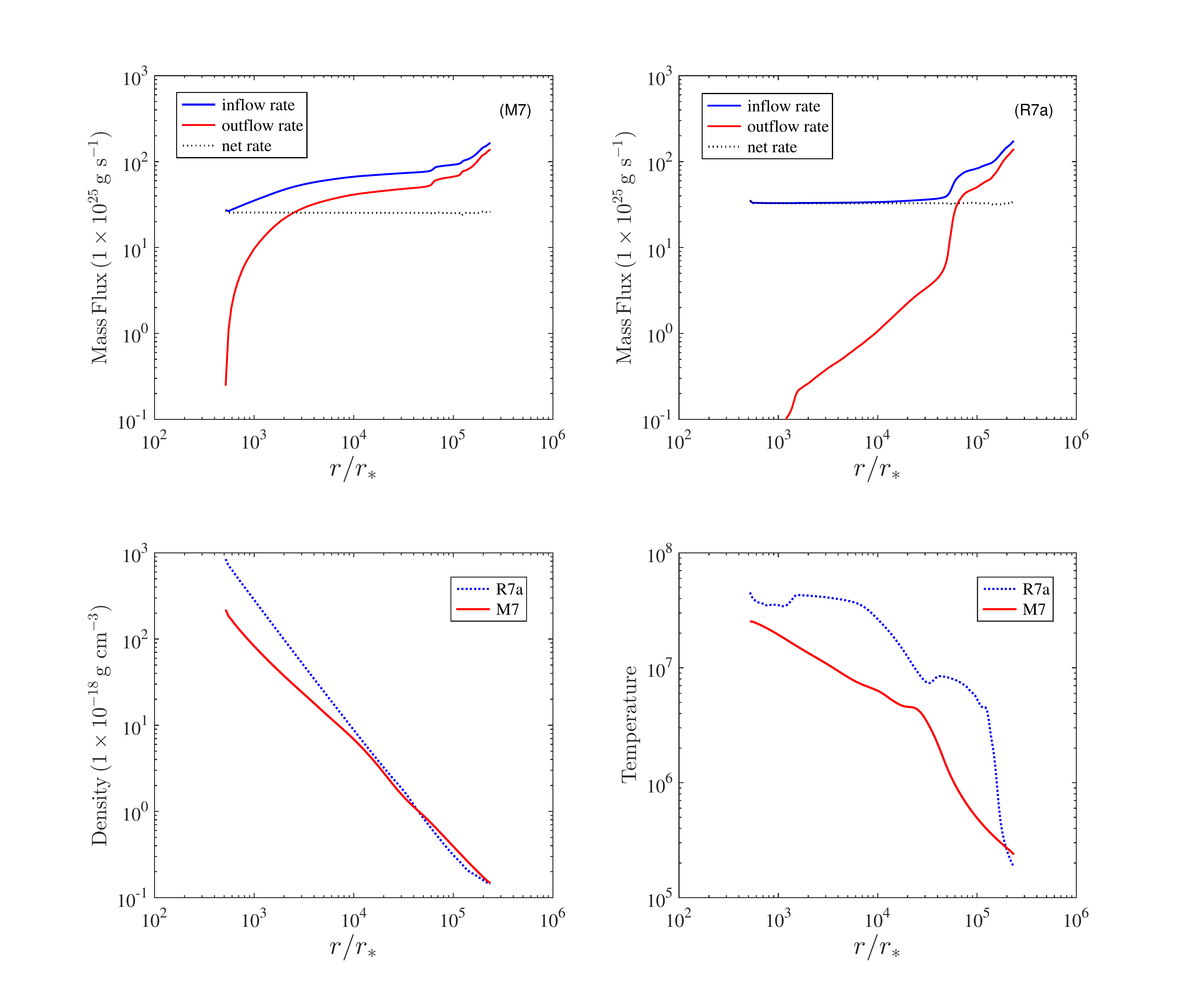}
 \caption{Comparison between models R7a and M7. The only difference between models M7 and R7a lies in the calculation of radiation force due to X-ray photons from the central AGN (see Section \ref{subsec:radiation_force}). Top panel: the radial profiles of mass inflow rate (blue solid lines), outflow rate (red solid lines) and the net rate (black dotted lines) in unit of $ 10^{25} \mathrm{g\, s^{-1} } $. Bottom panel: the radial profiles of gas density ({left panel}, in unit of $ 10^{-18} \mathrm{ g\, cm^{-3} } $) and temperature ({right panel}, in unit of $ \mathrm{K} $ ) averaged over three grids above the equatorial plane. The data are time-averaged.}
 \label{fig1}
\end{figure*}

We choose model R7a from \citet{Liu et al. 2013}  and compare it with its ``updated'' version, i.e., the fiducial model in our paper model M7. The parameters
are: $ M_\mathrm{BH} = 10^{8} M_{\sun} $, $ r_{*} = 3 r_\mathrm{s} = 8.8 \times 10^{13}\, \mathrm{cm} $,
$ L_\mathrm{Edd} = 1.25 \times 10^{46} \mathrm{erg\, s^{-1} } $, $ \rho_{0} = 5.0 \times 10^{-20}\, \mathrm{g} \,\mathrm{cm}^{-3} $,
and $ T_{0} = 2.0 \times 10^{6}\, \mathrm{K} $. The only difference between models M7 and R7a lies in the calculation of radiation force due to  X-ray photons from the central AGN, as we state in Section 3.4.

\begin{figure*}
 \includegraphics[width=\textwidth]{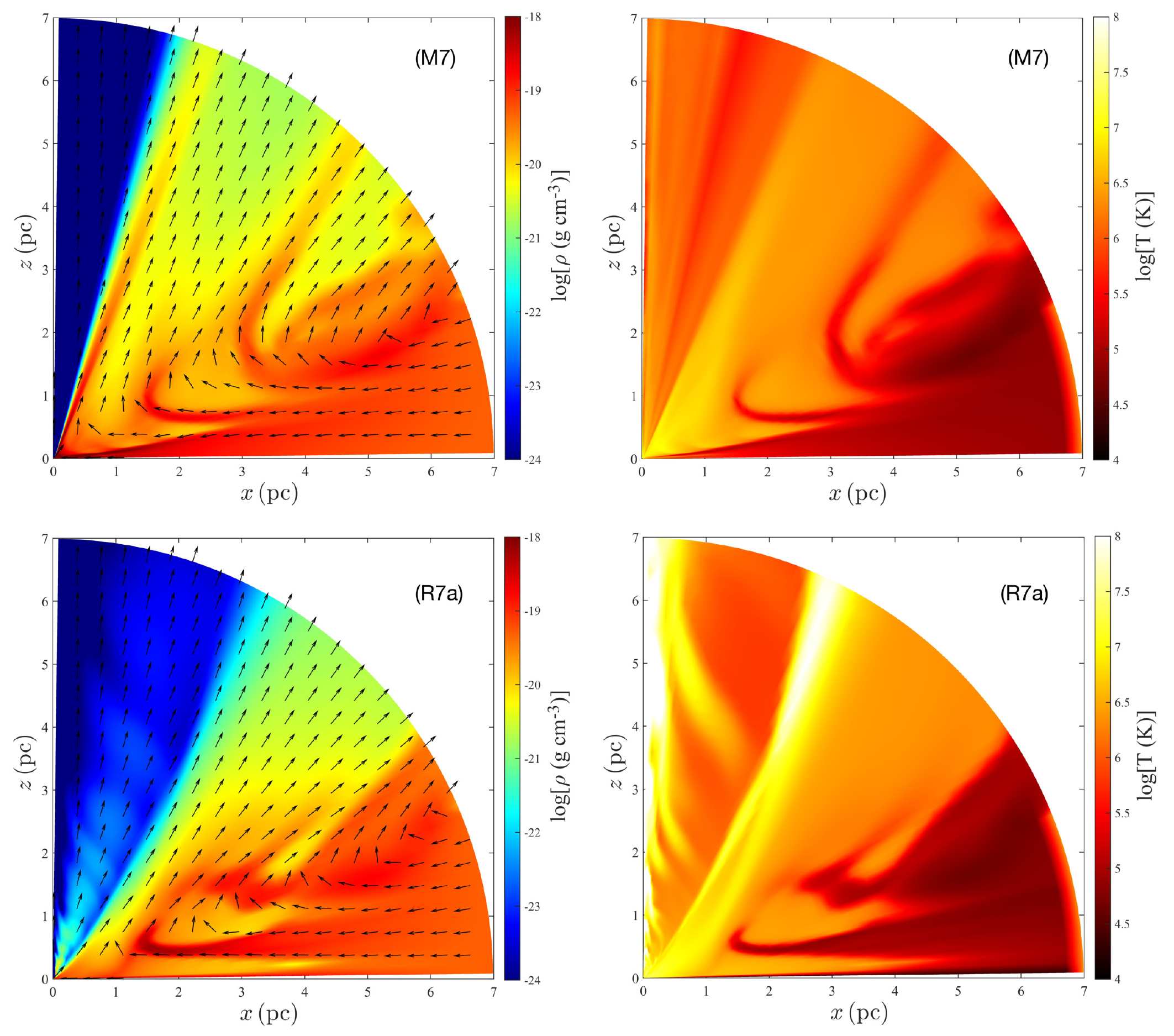}
 \caption{Snapshots of contours of logarithmic gas density (left column) and logarithmic temperature (right column) at the steady state. }
 \label{fig2}
\end{figure*}

Figure \ref{fig1} shows the comparison of some model results between the two models.
The top two panels show the radial profiles of the mass inflow, outflow and net rates.  The bottom two panels, from left to right,
show the radial profiles of gas density and temperature, respectively.
They are averaged over three grids above the equatorial plane.
The contours of the gas density and temperature of the two models
are shown in Figure \ref{fig2}.

Comparing the top-left and top-right panels of Figure \ref{fig1}, one can see that overall in model M7 the outflow rate becomes larger compared to model R7a, and  the inflow rate at the inner boundary of the simulation domain becomes smaller.
In model R7a, the inflow rate decreases from the outer boundary to $ \mathrm{6 \times 10^{4}} r_{*} $; while in model M7, the inflow rate keeps decreasing from the outer boundary to the inner boundary.
This is consistent with the fact that at small radii, the outflow rate in model M7 is larger than that in model R7a. For example, at $ \sim 2\times 10^{3} r_{\ast} $ the outflow rate in model R7a is
$ 0.2 \times 10^{25} {\rm g~s^{-1}}$, while it is $ 20 \times 10^{25} {\rm g~s^{-1}}$ in model M7.
The result is consistent with the left column of Figure \ref{fig2}. In this figure, we can see that compared to model R7a, the
dense outflow region can reach to a much smaller $\theta$ angle in model M7. In addition to the difference of mass flux between the two models, in model M7 the density and temperature of the gas around the equatorial plane  become lower compared to model R7a.

%From the bottom panel of Figure \ref{fig1}, we see that the midplane gas temperature in model M7 is significantly lower than that in model R7a. At small radii, the gas midplane density in model M7 is lower than that in model R7a. We can try to understand the reasons for the reduction of density from the top-left panel of Figure \ref{fig1} and from Figure \ref{fig2}. From these figures, we can see that in model M7 the inner region of the accretion flow is vertically expanded, i.e., the density scale height becomes higher in model M7. We speculate that this is because the stronger  the strong re-radiation force occurs in a wider $\theta$ range, i.e., $  20^{\circ} < \theta < 68^{\circ} $.
%The density scale height of the gas becomes slightly higher in model M7. Consequently, the inner edge of the accretion flow is vertically expanded and shields the radiation from the central engine that makes the accretion rate becomes
%higher in this model. The increase of the density in the region away from the midplane also causes the radiative cooling becomes stronger, which results in decrease of disc temperature (see the right column of Figure \ref{fig2}).
%Compare models R7a and M7, the disc morphology has not been changed too much with the modification of central X-ray radiation force (see Section 3.4).

\begin{figure*}
 \includegraphics[width=17cm]{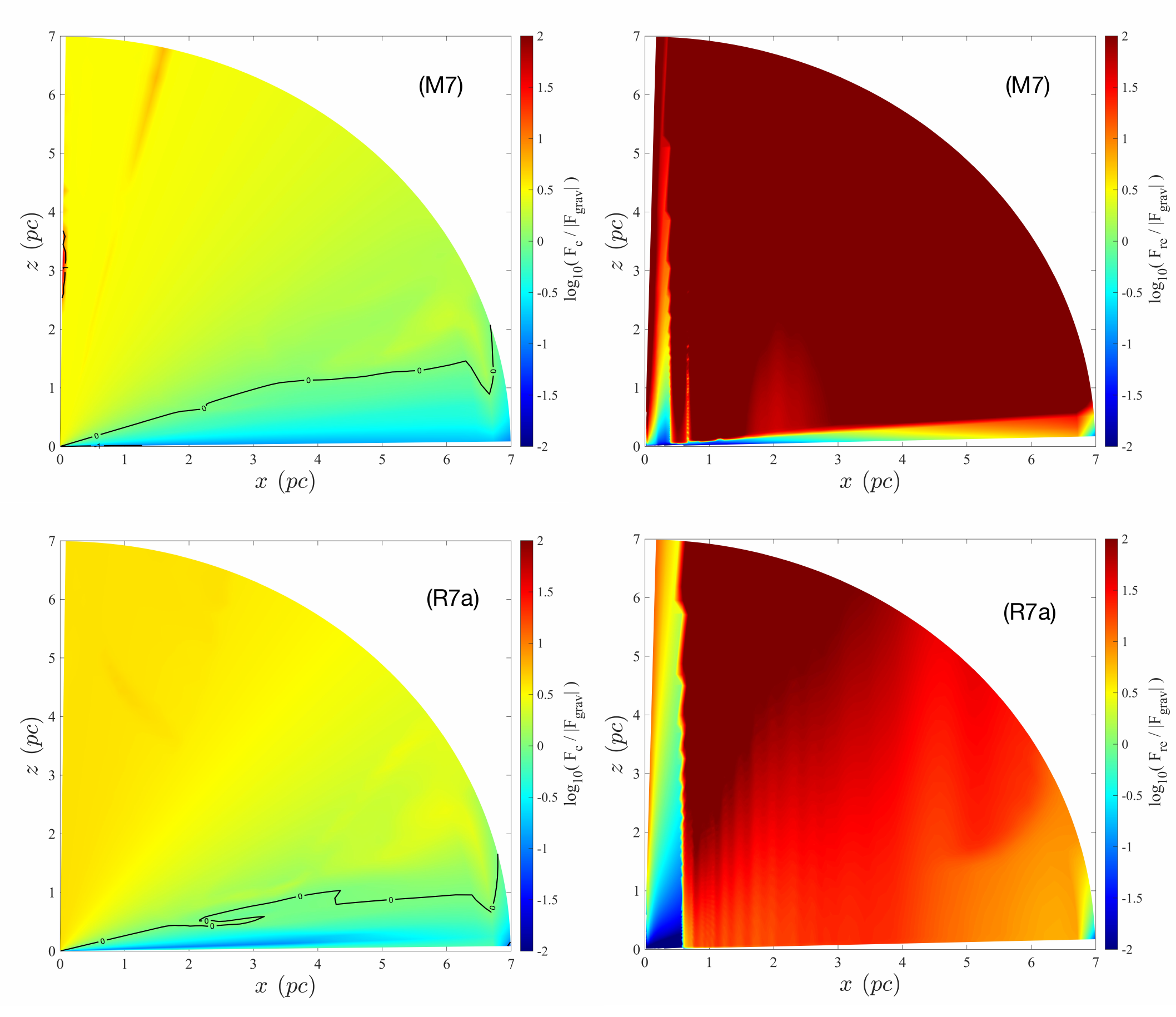}
 \caption{Comparison between total forces in radial ($ F_\mathrm{c} $) and vertical ($ F_\mathrm{re} $) directions for models R7a and M7. The forces are in units of the BH Newtonian gravity ($ F_\mathrm{grav} =  G M / r^2 $). The left panels represent the sum of the radiation force due to Thomson scattering of X-ray photons from the central AGN ($ F_\mathrm{c,X} $) and the line force due to the UV photons from the central AGN ($ F_\mathrm{c,o} $). The right panels show the re-radiation force in vertical direction due to Thomson scattering ($ F_\mathrm{re} $).The lines in the left column show the location where $ \log_{10} (F_\mathrm{c} / | F_\mathrm{grav} |) = 0 $. }
 \label{force_Fc_Fre}
\end{figure*}

To explain the above results, we plot the forces (normalized by gravity) in both radial and vertical directions in Figure \ref{force_Fc_Fre}.  This Figure has four panels. The top two panels are for model M7. The bottom two panels are for model R7a. In the left panels we plot the sum of the radiation force due to Thomson scattering of X-ray photons from the central AGN  ($ F_\mathrm{c,X} $) and the line force due to the UV photons from the central AGN ($ F_\mathrm{c,o} $). We note that the line force ($ F_\mathrm{c,o} $) is significantly stronger than $ F_\mathrm{c,X} $. In the right panels, we plot the re-radiation force in the vertical direction due to Thomson scattering ($ F_\mathrm{re} $). It is clear that the re-radiation force  ($ F_\mathrm{re} $) in the vertical direction is significantly stronger than the line force ($ F_\mathrm{c,o} $) in the radial direction. The re-radiation force can be decomposed into two forces, one is in the radial direction ($ F_\mathrm{re,r} $), the other one is in the theta direction ($ F_\mathrm{re, \theta} $). $ F_\mathrm{re,r} $ is significantly larger than the line force due to the UV photons from the central AGN ($ F_\mathrm{c,o} $). Therefore, in the present paper, the outflows are mainly driven by the re-radiation force. If the re-radiation force is not included, outflows can also be driven by the line force ($ F_\mathrm{c,o} $), but the mass flux of the outflow will be much smaller (see the comparison in Figure 1 of \citealt{Liu et al. 2013}).

From Figure. \ref{force_Fc_Fre}, we can see that the re-radiation force in the vertical direction in model M7 is much stronger than that in model R7a. Therefore, in model M7, gas can be more easily pushed to high latitudes. This is the reason that in model M7 the gas density at the mid-plane is lower. The gas density in the outflow region in model M7 is higher. Thus, the outflow mass flux in model M7 is higher. Because of the stronger mass loss in outflow in model M7, the radial density profile becomes flatter, thus the compression heating becomes weaker. This is why the temperature of the gas in model M7 becomes lower (see the bottom-right panel of Figure \ref{fig1} and Figure \ref{fig2}). From Eq. (\ref{fradz}), we can see that in the re-radiation force, there are three terms. We have calculated the individual terms and found that the largest term is the one related to the line cooling. In model M7, the gas density in outflow region is higher and temperature is lower. Therefore, the line cooling is stronger (see Eq. \ref{Lline}) in model M7, which induces a stronger re-radiation force.

%We can explain the above results as follows. In model M7, the line force due to the radiation directly from the central AGN becomes stronger, thus the velocity of the outflow in model M7 becomes larger (refer to the middle panel of Fig. 3). Consequently, because of the Bernoulli principle, more gas will be pushed out from the equatorial plane by the gas pressure. This is why the density of the gas in the equatorial plane in model M7 is lower (refer to the left-bottom panel of Fig. 1). Once this occurs, the re-radiation force will become stronger due to the larger density of the gas above the equatorial plane (refer to equation 23). This further makes the gas expand in the vertical direction so this is a kind of runaway instability (refer to Fig. 2). Because of the stronger mass loss in the outflow in model M7, the radial density profile becomes flatter, thus the compression heating with the accretion becomes weaker. This is why the temperature  of the gas in model M7 becomes smaller, as shown by the bottom-right panel of Fig. 1 and Fig. 2.

\begin{figure*}
 \includegraphics[width=16cm]{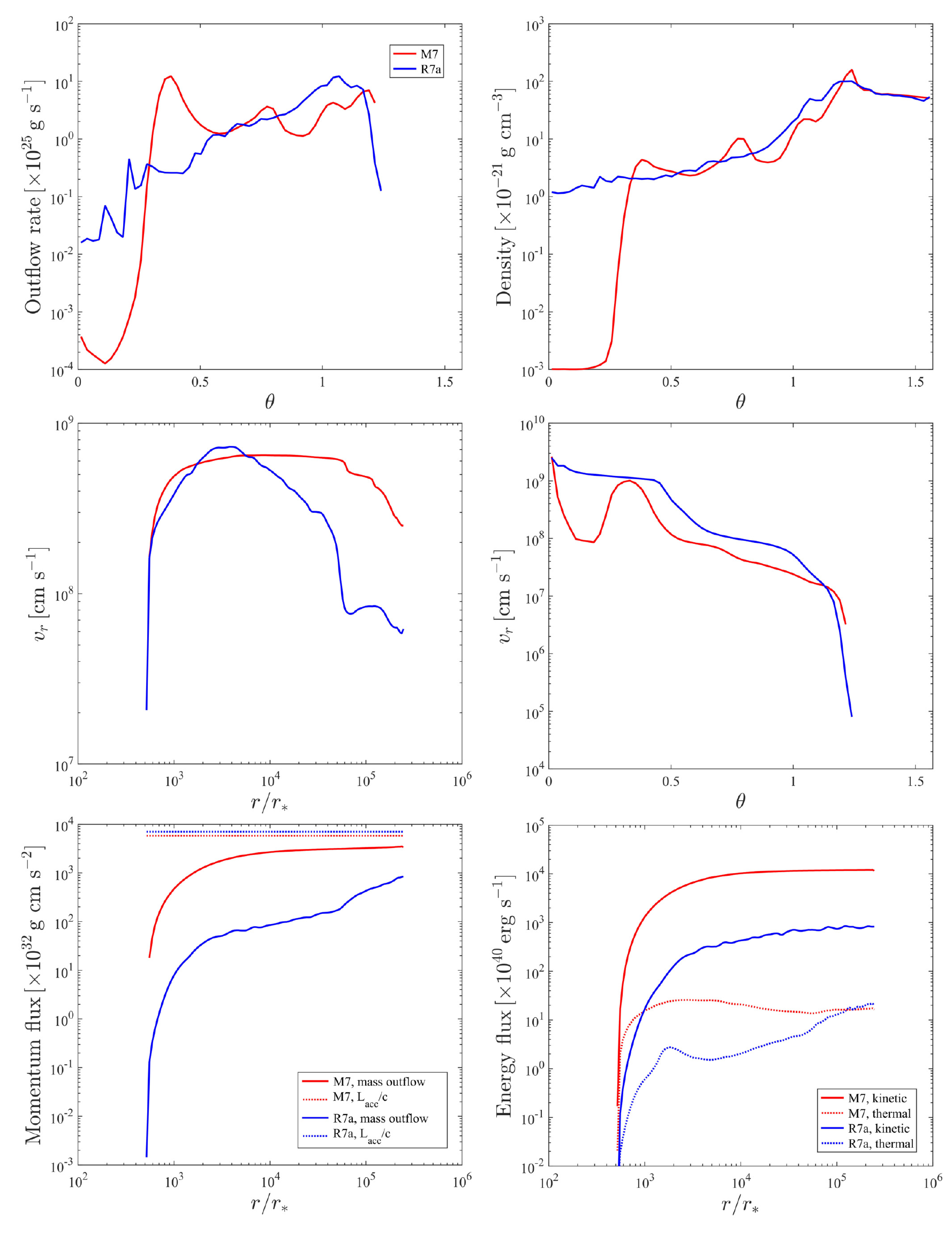}
 \caption{Properties of outflow in models M7 (red lines) and R7a (blue lines).
Top: the angular distribution of the mass outflow rate (top-left panel, in unit of $ 10^{25} \mathrm{g\, s^{-1} } $; defined as $\dot{M}_{\rm out}(\theta)=2\pi r^2\rho ~{\rm max}(v_r,0)sin(\theta) \Delta \theta$)
and density (top-right panel, in unit of $ 10^{-21} \mathrm{g\, s^{-1} } $ at the outer boundary of the simulation domain; Middle: the radial profile of the mass flux-weighted radial velocity of the outflow (left panel, in unit of $ \mathrm{cm\, s^{-1}} $) and angular distribution of the outflow velocity (right panel, in unit of $ \mathrm{cm\, s^{-1}} $);
Bottom-left: the radial profile of the momentum flux of the outflow (in unit of $ \mathrm{10^{32}\, g\, cm\, s^{-2}} $;  solid lines) and
radiation flux from the central AGN ($ \mathrm{L_{acc}/c} $;  dotted lines).
Bottom-right: the radial profile of the kinetic (the solid lines) and
the thermal (dotted lines) energy fluxes in unit of $ \mathrm{10^{40}\, ergs\, s^{-1}} $.
All the lines in this figure are time-averaged.}
 \label{fig3}
\end{figure*}

Additional properties of the outflow for the fiducial runs have been shown in Figure \ref{fig3}.
In this figure, similar to the Figure 5 in \citet{Liu et al. 2013},  we show the time-averaged angular and radial profiles of some outflow properties. Red and
blue colours represent M7 and R7a models, respectively. The first row, from left to right, shows the angular distribution of the mass flux and density at the outer boundary of the simulation domain,
i.e, $ \sim\,  6 \mathrm{pc} $. We can see that no outflow is present in the range of $ \theta >  68^{\circ}$. Compared to model R7a, the main mass flux of the outflow in M7 occurs in a broader range of angles, $ \mathrm{20^{\circ} < \theta < 68^{\circ}} $. The time-averaged angular profiles of  density for the two models are similar.  The middle-left panel shows the radial profile of the mass flux-weighted radial velocities of the outflows. We can see from the figure that  the velocity increases sharply with radius in the innermost  region. The velocity reaches its maximum values around $\sim 10^3 r_*$. At  larger radii, the velocity decreases with increasing radius.
Such a decrease  is mainly due to the injection of new gas into the outflow
at large radius. We note that in model M7 the decrease of  velocity at large radii occurs only until $6\times 10^4 r_*$, while in model R7a it occurs at a much smaller radius, $4\times 10^3r_*$. Overall, the velocity of outflow in model M7 is significantly larger than that in model R7a. This is because of the enhancement of the radiation force in model M7. The middle-right panel shows the angular distribution of outflow velocity at $ 6 \, \mathrm{pc} $. We can see a similar behaviour in two models except for a sudden drop near the polar region.

 We define fluxes of momentum, kinetic energy, and thermal energy of outflow as follows,
\begin{equation} \label{p_dot_w}
	\dot{p}_{w} (r) = 4 \pi r^{2} \int^{\pi/2}_{0} \rho v_{r}^{2} \sin(\theta)\, \mathrm{d} \theta \qquad \mathrm{for}\, \, v_{r} > 0,
\end{equation}
\begin{equation} \label{E_dot_k}
	\dot{E}_{k}(r) = 2 \pi r^{2} \int^{\pi/2}_{0} \rho v_{r}^{3} \sin(\theta)\, \mathrm{d} \theta \qquad \mathrm{for}\, \, v_{r} > 0,
\end{equation}
\begin{equation} \label{E_dot_th}
	\dot{E}_\mathrm{th}(r) = 4 \pi r^{2} \int^{\pi/2}_{0} e v_{r} \sin(\theta)\, \mathrm{d} \theta \qquad \mathrm{for}\, \, v_{r} > 0,
\end{equation}
The solid and dotted lines in the bottom-left panel of Figure \ref{fig3} show the radial profiles of  these quantities in both models. In accordance with the mass flux and velocity of wind in the two models shown above, the momentum flux of wind in model M7 is much larger than that in model R7a. Also shown in the figure is the momentum flux of radiation from the AGN for comparison purpose. It can be seen that the momentum flux of radiation
is higher than that of  outflow at all radii. This is reasonable. In both models,
the momentum flux of wind, $ \dot{p}_{w} $, increases with increasing radius
and gets closer to the radiation flux with the increasing radius.  The percentage of the radiation momentum flux entrained by the wind is even as high as  $\sim 60\%$ for model M7. This value is much higher than that of  model R7a. The reason is that, as we can see from Figure \ref{fig2}, the density of the gas in the wind region in model M7 is higher than that of model R7a so more radiation momentum can be captured by wind in model M7.
The bottom-right panel of Figure \ref{fig3} shows the kinetic
($ \dot{E}_k $) and thermal energy fluxes ($ \dot{E}_\mathrm{th} $)
of the outflow. The kinetic energy of outflow in model M7 is about 10 times higher than that in model R7a. In both models, the kinetic energy flux is much higher than the thermal energy flux so the wind motion is supersonic.

\begin{figure*}
 \includegraphics[width=\textwidth]{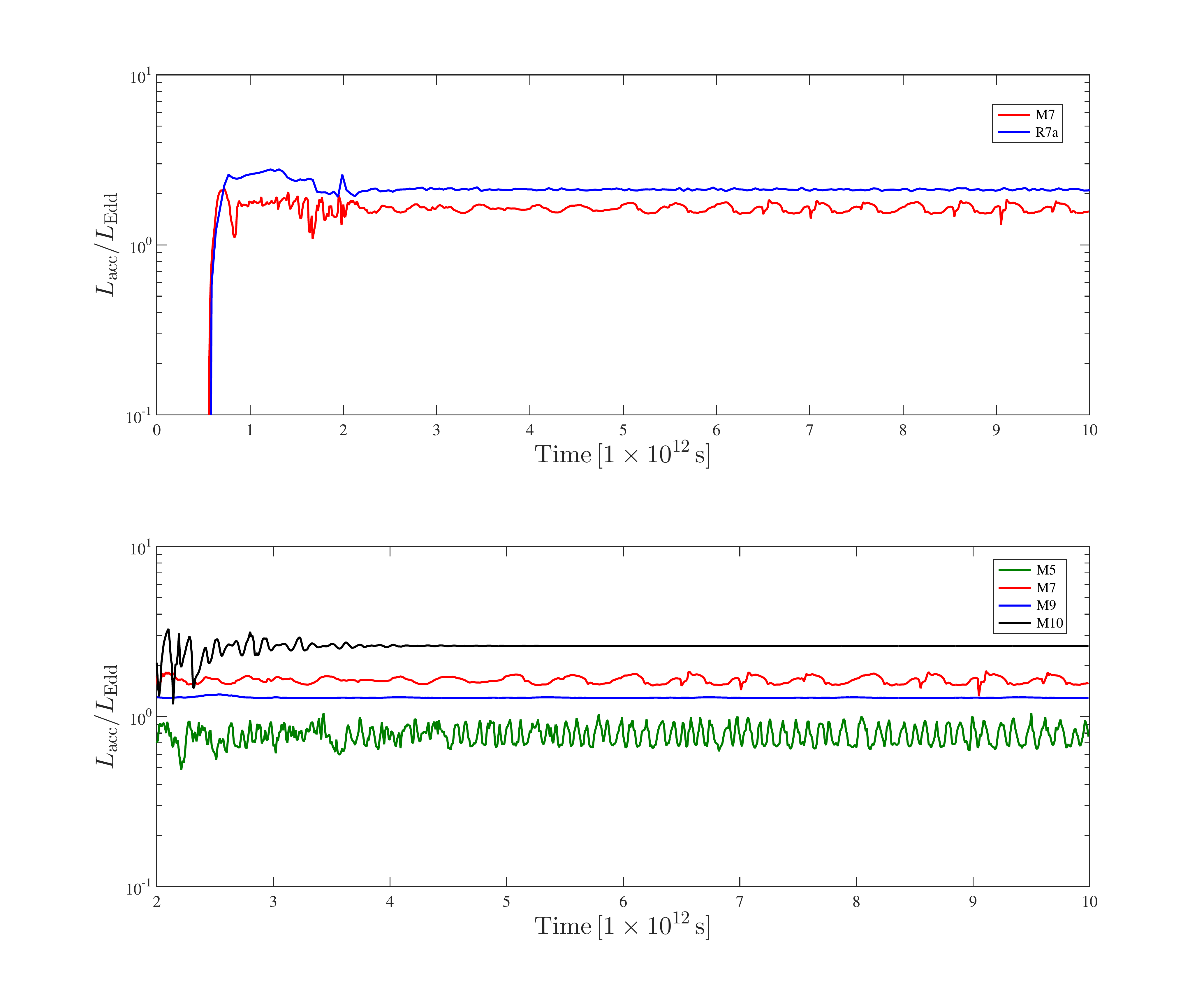}
 \caption{Time evolution of the AGN luminosity in unit of Eddington luminosity $ L_\mathrm{Edd} $ for different models. Top panel: comparison between models R7a and M7. Bottom panel: The green, red, blue and black lines correspond to models M5, M7, M9 and M10, respectively.}
 \label{fig4}
\end{figure*}

The top panel of Figure. \ref{fig4} shows the time evolution of AGN luminosity in unit of Eddington luminosity, $ L_\text{acc}/ L_\text{Edd} $ for models M7 and R7a.
Column 5 in Table \ref{table1} also shows the corresponding time-averaged luminosities with the standard deviations $ \sigma_{n} $. One problem \citealt{Liu et al. 2013} hoped to solve by investigating the line-force driven wind is the so-called sub-Eddington puzzle. That is, observations show that the luminosity of almost all
AGNs are sub-Eddington (\citealt{Kollmeier et al. 2006}; \citealt{Steinhardt and Elvis 2010}), while theoretically the luminosity of an accretion flow can easily be
super-Eddington given the abundant gas supply in galactic center region. This puzzle is not solved in \citealt{Liu et al. 2013}, i.e, the wind is not strong enough to reduce the AGN accretion rate. Although the line force is increased in the present work,  the luminosity in model M7 is still super-Eddington, $ L_\text{acc}/ L_\text{Edd}  \sim 1.64 $. Thus the sub-Eddington puzzle has not be solved by these modifications. One reason may be that the accretion rate should continue to decrease within the inner boundary of our simulation domain. Another reason is that in addition to radiation which we have included here, the AGN will also produce winds  and these winds will also interact with the gas in our simulation domain and reduce the accretion rate (e.g., \citealt{Ciotti and Ostriker 1997, Ciotti and Ostriker 2001, Ciotti and Ostriker 2007}; \citealt{Yuan et al. 2018}; \citealt{Bu and Yang 2019}).

\subsection{The effects of gas temperature and density}

\begin{figure*}
 \includegraphics[width=\textwidth]{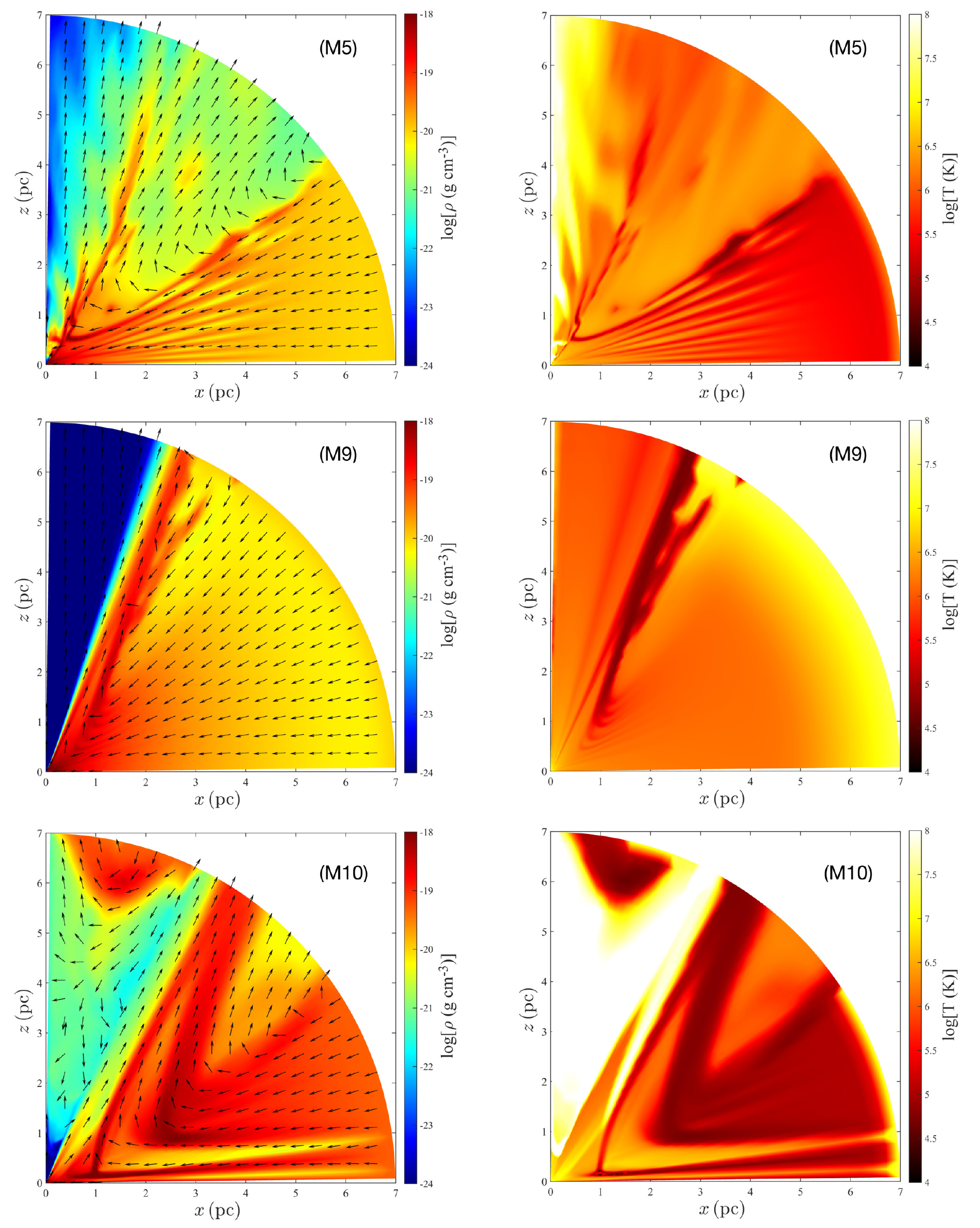}
 \caption{Snapshots of contours of logarithmic gas density (left-hand column) and
logarithmic temperature (right-hand column) correspond to models M5, M9 and M10,
respectively. The density over-plotted with the directions of poloidal velocity as arrows
(left column). The figures are placed in order of increasing density  and tempruture at
$ r = r_\mathrm{max} $ (see Table \ref{table1}).}
\label{fig5}
\end{figure*}

We study the effect of $ T_0 $ and $ \rho_0 $ in this section.  The corresponding models are presented
in Table. \ref{table1}.  The bottom panel of Figure \ref{fig4} shows the time evolution of the accretion luminosity for models M5 (green line), M7 (red line), M9 (blue line), and
M10 (black line).  Figure \ref{fig5} shows the snapshots of contours of gas density and temperature of models  M5, M9 and M10.

From Figure \ref{fig4}, we see that with the increase of $ \rho_{0} $, the accretion luminosity increases. This is not surprising. What is interesting is the effect of $ T_0 $. From model M5 to M9 and from M7 to M10, the only change is the increase of $T_0$. We can see from Figure \ref{fig4} that the increase of $ T_0 $ has two effects. One is the increase of luminosity; another one is that the light curve becomes very smooth. These two effects are related with each other, as we explain below. To explain such effects, the key point is that the value of the line force is a function of gas temperature (see \citealt{Proga et al. 2000} for details).  When $T_0$ increases from $2\times 10^6{\rm K}$ to $2\times 10^7{\rm K}$, the line force significantly decreases. Consequently, as we can see from Figure \ref{fig5},  the inflow region becomes much wider, from $\theta\sim 60^{\circ}-90^{\circ}$ in model M5 to $\theta \sim 30^{\circ}-90^{\circ}$ in model M9. This explains why the accretion rate and subsequently the luminosity increases. In model M9, when the accreting gas reaches the innermost region, its density becomes very large and temperature significantly decreases because of the strong radiative cooling. Radiative line force then begins to play its role and produce an outflow. This is the origin of the high-density filament-like outflowing structure seen in model M9 at $\theta\sim 30^{\rm \circ}$.

 In model M5, the  temperature in all $\theta$ region is not high so the line force can play its role in all these region. Consequently, when luminosity becomes strong the line force also becomes strong; thus accretion can be suppressed and luminosity decreases. This results in a weaker line force and subsequently stronger accretion and luminosity. This explains why we see strong fluctuation in the light curve of model M5. Compared to model M5, in model M9, the difference of temperature of inflow and outflow is significant and the two regions are spatially well separated. The temperature in the inflow region is always high, while the temperature in the outflow region is always low. So there is almost no fluctuation in this model and the light curve  is smooth.

\subsection{Observational implications}

Outflow is widely observed in different kinds of AGNs, usually in the form of blue-shifted absorption lines (BAL) (e.g., \citealt{Crenshaw et al. 2003}; \citealt{Tombesi et al. 2010, Tombesi et al. 2012a, Tombesi et al. 2012b}; \citealt{Kaastra et al. 2012}; \citealt{Gofford et al. 2015}; \citealt{He et al. 2019}). In \citealt{Gofford et al. 2015}, they performed a systematic analysis to the outflow properties in a sample of 51 {\it Sukaku}-observed AGNs. They found that the properties of wind cover a large range. Typically they are detected at distance from $\sim 0.001-1$ pc, the outflow velocity is in the range between $0.01-0.1$c, the mass outflow rate falls between $10^{24}-10^{26}{\rm g~s^{-1}}$, and the kinetic luminosity of wind ranges between $10^{43}-10^{45}{\rm erg~s^{-1}}$.

We now examine whether the observed wind properties are consistent with our theoretical predictions. We choose  M5 since its corresponding accretion luminosity is $79\%L_{\rm Edd}$, which is close to the typical luminosity of a luminous AGN as sampled in \citealt{Gofford et al. 2015}. We find that the typical  outflow velocity is $ 0.02 c $, the outflow mass flux becomes significant from $ 0.07 $pc, the maximum mass outflow rate achieved at the outer boundary is  $\sim 4.7 \times 10^{26}{\rm g~s^{-1}}$ (Table \ref{table1}), and the kinetic luminosity of outflow is  $\sim 3\times 10^{43}{\rm erg~s^{-1}}$. All these values are roughly in the observed range of outflow properties.
  
As we explained in subsection \ref{the_fiducial_run}, the outflow found in this paper is driven by both the line force and re-radiation force. The properties of the outflows are consistent with observations (\citealt{Gofford et al. 2015}). Therefore, the BAL winds in AGNs may be driven by the combination of line and re-radiation forces. As mentioned above, if re-radiation force is neglected, outflow can be driven by line force. However, in this case, the mass flux of outflow is lower. Therefore, we cannot eliminate the possibility that the observed BAL outflow is driven by line force. We can then conclude that with the help of re-radiation force, the BAL outflow can be stronger.

\section{Summary and Discussion} \label{sec:Summary_Discussion}

In this paper, we have investigated the dynamics of  accretion flow with the irradiation from the central AGN by performing two-dimensional hydrodynamical simulations. We focus on relatively large radii, with the simulation domain ranging from $ 0.01-7\, \mathrm{pc} $. The mass of the central black hole is  $ M_\mathrm{BH} = 10^{8} M_{\odot} $.  The central engine has two components, i.e., a standard thin disc and a spherical hot corona, which emits optical/UV and X-ray radiation. Both the optical/UV and X-ray photons can produce a radiation force by Thomson scattering. Moreover, the optical/UV photons can also produce a line force, since the gas is not fully ionized thus the interaction cross-section is significantly amplified compared to the pure Thomson scattering. The main improvement in our present work compared to our previous one (i.e., \citealt{Liu et al. 2013}) is that we now correctly include the radiation force corresponding to the Thomson scattering by the X-ray photons (equation \ref{F_cent} in section \ref{subsec:radiation_force}).

Taking a fiducial model M7 as an example, we find that such an update of the radiation force results in an enhancement of the mass outflow flux compared to the previous corresponding model (i.e., model R7a in \citealt{Liu et al. 2013}) (Fig. \ref{fig1}). We have calculated the total forces in the radial and vertical directions in Fig. \ref{force_Fc_Fre}. we find that the re-radiation force in vertical direction in model M7 is much stronger than that in model R7a. Correspondingly, the mass inflow rate decreases, and the temperature and density of the gas around the equatorial plane also change somewhat (Fig. \ref{fig1}). The typical velocity of outflow also increases in model M7 in most region  by a factor of a few  (middle panel of Fig. \ref{fig3}). Because of the increase of mass outflow rate and velocity, the momentum flux and kinetic flux of outflow increases roughly by a factor of 10 (bottom panel of Fig. \ref{fig3}). For model M7, the momentum flux of outflow is as high as $60\%$ of the momentum flux of radiation. This percentage is significantly higher than model R7a. This is because the density of outflow in model M7 is higher thus can entrain the momentum of more photons. 

We have also examined the effects of density and temperature of the gas in the initial state of our simulations, $\rho_0$ and $T_0$. We find that the effect of $T_0$ is interesting. When $T_0$ becomes higher, the inflow rate increases (Fig. \ref{fig4}). This is because the line force becomes weaker due to the increase of $T_0$ thus accretion becomes stronger. Moreover, when this gas is accreted into the innermost region, the density becomes high and thus temperature decreases (Fig. \ref{fig5}) so the line force becomes strong; a strong outflow will be produced there. Such an outflow has a filamentary-like structure (Fig. \ref{fig5}) and the outflow region is well separated from the inflow region. This results in a smooth light curve of the accretion luminosity, rather than the fluctuated light curve when $T_0$ is lower (bottom of Fig. \ref{fig4}).

We have also compared the properties of outflow obtained in our simulations with those outflow obtained from blue-shifted absorption line observations (\citealt{Gofford et al. 2015}). These properties include the mass outflow rate, velocity, and kinetic luminosity at a given distance from the central black hole. While the observed range is large, we find that our simulation results are well within the observed range. Thus the combination of radiation line-force and re-radiation force we have studied in this work could be a candidate mechanism  for the widely observed outflow in AGNs.

There are several caveats in this study. One is that our
 approach for calculating the radiation and re-radiation forces is rather simple. In principle, the full
radiation transfer calculation is desirable. Simulations with radiative transfer are expensive but more realistic.
Another caveat here is that, we do not include the interaction of radiation with dust.
The photon-dust interaction can enhance the mass flux of outflow. In the present work, we only consider the low-angular momentum case. Next step, we will consider the case of a large-angular momentum.  In this case, a rotationally supported disc will be formed and the dynamics of accretion flow may change. Another important case one may consider is the effect of magnetic field. Usually the inclusion of magnetic field will enhance the outflow significantly. It will be interesting to study outflow by combining the radiation and magnetic field.

\section*{Acknowledgements}

 Amin Mosallanezhad is supported by the Chinese Academy of Sciences
President's International Fellowship Initiative, (PIFI), Grant No. 2018PM0046.
D.-F.B. is supported in part by the Natural Science Foundation
of China (grant 11773053).
This work is supported in part by the National Key Research and Development Program of China (Grant
No. 2016YFA0400704), the Natural Science Foundation of China (grants 11573051, 11633006, 11650110427,
11661161012, U1431228, 11233003, and 11421303), the Key Research Program of 
Frontier Sciences of CAS (No. QYZDJSSW-SYS008), and
the Astronomical Big Data Joint Research Center co-
founded by the National Astronomical Observatories,
Chinese Academy of Sciences and the Alibaba Cloud.
The computation has made use of the High Performance Computing 
Resource in the Core Facility for Advanced Research
Computing at Shanghai Astronomical Observatory.

\end{document}